\documentclass[prd,aps,twocolumn,cite,superscriptaddress,showpacs,nofootinbib, amsmath, amssymb,preprintnumbers]{revtex4-2}

\pdfoutput=1
\usepackage{graphicx}
\usepackage{amsmath, amssymb}
\usepackage{slashed}
\usepackage{xcolor}
\usepackage{array,arydshln}
\usepackage{enumerate}
\usepackage[italicdiff]{physics}
\usepackage[pdftex,
  colorlinks=true,
  citecolor=green!60!blue]{hyperref}
\usepackage{url}

\begin{document}

\preprint{KANAZAWA-26-04}
\preprint{RIKEN-iTHEMS-Report-26}

\title{A lower bound on primordial power spectrum from halo substructure}

\author{Shin'ichiro~Ando}
\affiliation{GRAPPA Institute, University of Amsterdam, 1098, XH Amsterdam, The Netherlands}
\affiliation{Kavli Institute for the Physics and Mathematics of the Universe, University of Tokyo, Chiba 277-8583, Japan}

\author{Nagisa~Hiroshima }
\affiliation{Department of Physics, Faculty of Engineering Science, Yokohama National University, Yokohama 240–8501, Japan}
\affiliation{RIKEN Interdisciplinary Theoretical and Mathematical Sciences (iTHEMS),
Wako, Saitama 351-0198, Japan}

\author{Koji~Ishiwata}
\affiliation{Institute for Theoretical Physics, Kanazawa University, Kanazawa 920-1192, Japan}

\author{Tomo~Takahashi}
\affiliation{Department of Physics, Saga University, Saga 840-8502, Japan}

\begin{abstract}

  We investigate how the primordial curvature perturbation of
  a certain wavelength scale affects the halo and subhalo
  structure. 
  Primordial power spectrum considered in this paper features a nearly scale-invariant form with a cutoff at the wavenumber $ k = \order{1}~{\rm Mpc}^{-1}$ and an additional log-normal bump at smaller scales.
  We compute the host halo evolution, as well as the
  subhalo mass function. To be consistent with the
  observations of stellar streams and gravitational lensing data, the amplitude of the bump that is typically the
  same or larger than $10^{-9}$ of the primordial curvature
  perturbation should be present at the wavenumbers of
  $10$\,--\,$30$~Mpc$^{-1}$.

\end{abstract}

\maketitle

\section{Introduction}

Inflation is a paradigm of the early universe, which is strongly
supported by
various cosmological observations such as cosmic microwave background (CMB), large scale structure and so on. 
During inflation, the vacuum energy is dominated by a
slow-rolling inflaton field and the quantum fluctuations of the
inflaton generate the primordial curvature perturbation, which plays
a role of a seed for the structure formation in matter-dominated
universe. Though it is 
strictly constrained to $\sim 10^{-9}$ on
large scales, the amplitude of the primordial curvature perturbation 
is less constrained on smaller scales, 
$k\gtrsim \order{1}\,{\rm Mpc}^{-1}$.

Recently, the observations of galactic structures
such as satellite counts, stellar streams, which are believed to be remnants of disrupted satellite galaxies, and gravitational lensing reveal the 
small-scale structures of the halo of  
the Galaxy~\cite{DES:2019vzn,Hezaveh:2016ltk,Banik:2019cza,Banik:2019smi}. 
The evolution of hierarchical structures of halos,
on the other hand, is investigated from both the numerical and analytical ways. N-body simulations are powerful in probing detailed physical processes, while
analytic or semi-analytic schemes are powerful in covering a wide range of mass and redshift of interest. Press-Schechter or extended Press-Schechter (EPS) formalism is a representative of such strategies. 
\texttt{SASHIMI}\footnote{\url{https://github.com/shinichiroando/sashimi-c}}~\cite{Hiroshima:2018kfv,Ando:2019xlm} is a
python package for the evolution and properties of the subhalos. 
It incorporates an analytic description of tidal evolution of subhalos~\cite{Jiang:2014nsa} in the EPS
formalism.
With the code, the primordial curvature perturbation and structure of the
subhalos are  
connected. With the observations mentioned above, therefore, we can get better knowledge about the curvature perturbation of the smaller scale.
Specifically, 
the fact that we observe some structure on small scales 
indicates that there should be a {\it lower} bound on primordial fluctuations to source those structure.

In this paper, we derive a {\it lower} bound on the primordial power spectrum 
on small scales
from observed halo substructure.
To this end, we model
the primordial power spectrum of the curvature perturbation ${\cal P}_\zeta$ that has
a cutoff and a log-normal bump.
More precisely,
the nearly scale-invariant power
spectrum has a cutoff at $k=k_{\rm cut}$ and an additional log-normal
bump in $k> k_{\rm cut}$,
where we take $k_{\rm cut} = 5~h\,{\rm Mpc}^{-1}$ in our analysis.
This modeling is motivated by the fact that,  on large scales, the primordial power spectrum is well constrained by observations of CMB, large scale structure, including Lyman-$\alpha$ data up to $k \sim {\cal O}(1)\,{\rm Mpc}^{-1}$.\footnote{
From the perspective of model-building, primordial power spectrum featuring a cutoff,  
below which its amplitude is significantly suppressed, has been discussed in the context of broken-scale-invariance model~\cite{Starobinsky:1992ts,Kamionkowski:1999vp}, double inflation~\cite{Yokoyama:2000tz,Kubota:2022pit}, thermal inflation~\cite{Hong:2015oqa,Bae:2022gkv}, inflaton-spectator model~\cite{Enqvist:2019jkb}, etc. 
}
On the other hand, measurements of the small-scale power spectrum are still limited  
and only weak constraints have been obtained from CMB spectral distortions \cite{Chluba:2012we},  ultracompact minihalos \cite{Bringmann:2011ut}, primordial black holes  \cite{Byrnes:2018txb}, and so on, all of which provide an {\it upper} bound on the primordial power spectrum.
In contrast,  we in this paper aim to place a {\it lower} bound on ${\cal P}_\zeta$ from observed substructure, which gives important implications for the generation mechanism of primordial fluctuations. 
Although we consider a power spectrum with a cutoff to analyze a lower bound on $P_\zeta$ on small scales, models that suppress small-scale fluctuations, for example through a negative spectral running, would also be constrained by the lower bound derived in this paper.

Given the primordial power spectrum of the curvature perturbation, we compute the time evolution of the host halo mass and
the subhalo mass distribution function in the host halo, and analyze
the consistency with the observational data.  In the past work, a
similar bumpy power spectrum without the cutoff was
considered~\cite{Ando:2022tpj}.  With the setup, the halo mass
evolution occurs in a continuum way due to non-zero curvature
perturbatioin at any scale. In the present study, on the contrary, 
we analyze the impact of zero and nonzero power spectrum of the curvature perturbation on the halo and subhalos evolution,
which 
aims at deriving 
lower bounds on the primordial power spectrum.
Our present work includes some
improvements from Ref.\,\cite{Ando:2022tpj}. First, we take into
account the observational data from the stellar streams and
gravitational lensing, which gives a direct constraint on the subhalo
mass distribution function. Second, we compute the host halo mass
evolution in a new way. It gives a consistent result with the numerical
approach adopted in Ref.\,\cite{Hiroshima:2022khy}, 
which reduced the computational cost significantly and boosts the efficiency of the overall calculation.

In our study, we use the cosmological parameters based on the Planck
2018 results~\cite{Aghanim:2018eyx}; the scalar amplitude
$A_s=2.092\times 10^{-9}$, the scalar spectral index, $n_s=0.965$, the
pivot scale $k_*=0.05~{\rm Mpc}^{-1}$, the energy density of dark
matter and baryon $\Omega_Mh^2=\Omega_{\rm
  DM}h^2+\Omega_bh^2=0.12+0.02237$, and $h=0.6736$.

\section{The model of curvature perturbation}
\label{sec:Pzeta}

\begin{figure}[t]
  \begin{center}
    \includegraphics[scale=0.55]{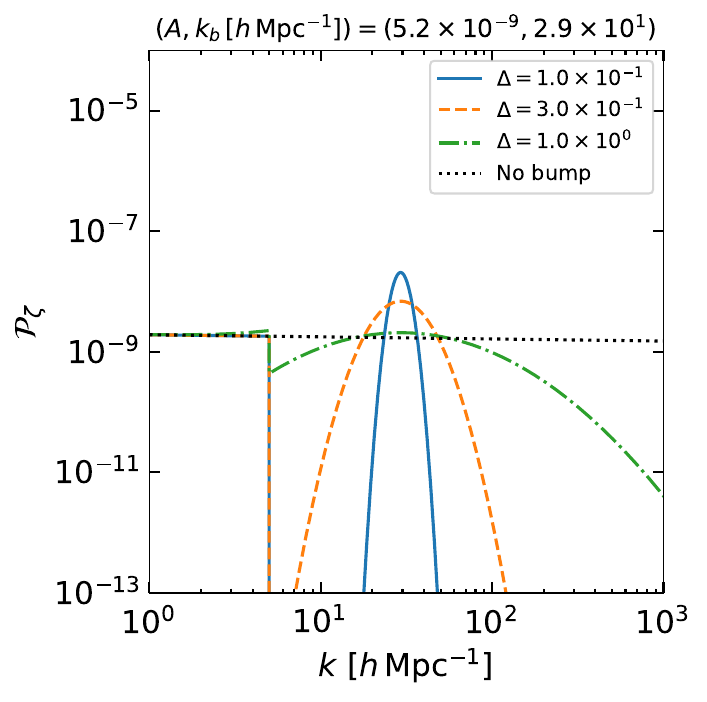}
  \end{center}
  \caption{\small 
    Primordial curvature power spectrum ${\cal P}_\zeta$ modeled as Eq.~\eqref{eq:P_zeta_bump}
    as a function of
    wavenumber. $k_b\,[h\,{\rm Mpc}^{-1}]=29$ 
    and $A=5.2\times 10^{-9}$ are taken. The value of $\Delta$  is indicated in
    the figure. As a reference, 
    the power spectrum without a cutoff
    ${\cal P}_\zeta^{\rm CMB}$ is plotted as `No bump.'}
  \label{fig:Pzeta}
\end{figure}

\begin{figure}[t]
  \begin{center}
    \includegraphics[scale=0.55]{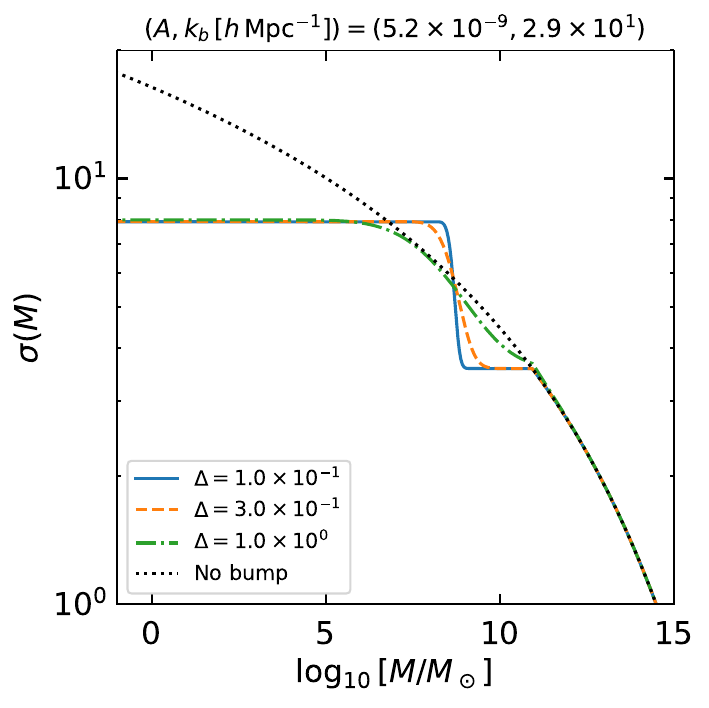}
  \end{center}
  \caption{\small Square root of the variance of the matter power spectrum. The
    parameters are the same as Fig.\,\ref{fig:Pzeta}. As in the
    figure, the variance calculated by ${\cal P}_\zeta^{\rm CMB}$ is
    shown. }
  \label{fig:sigma}
\end{figure}

In the following, we study the impact on the subhalo mass distribution to derive a lower bound on the primordial power spectrum. For this purpose,
we consider a model in which the primordial power spectrum has a cutoff and a bump:
\begin{align}
\label{eq:P_zeta_bump}
  {\cal P}_\zeta = {\cal P}_\zeta^{\rm CMB}\Theta(k_{\rm cut}-k)
  +{\cal P}_\zeta^{\rm bump}\,,
\end{align}
where $\Theta(x)$ is the Heaviside step function and 
\begin{align}
  {\cal P}_\zeta^{\rm CMB} &= A_s\left(\frac{k}{k_*}\right)^{n_s-1}\,,\\
  {\cal P}_\zeta^{\rm bump}&=\frac{A}{\sqrt{2\pi}\Delta}
  \exp \left[-\frac{\ln^2(k/k_b)}{2\Delta^2}\right]\,.
\end{align}
Here 
$k_{\rm cut}$, $A$, and $\Delta$ are constants and we take $k_{\rm
  cut}=5~h\,{\rm Mpc}^{-1}$ throughout this
paper. 
Fig.\,\ref{fig:Pzeta} is an example of the curvature
perturbation with this modeling 
for several values of $\Delta$.

\begin{figure*}[t]
  \begin{center}
    \includegraphics[scale=0.45]{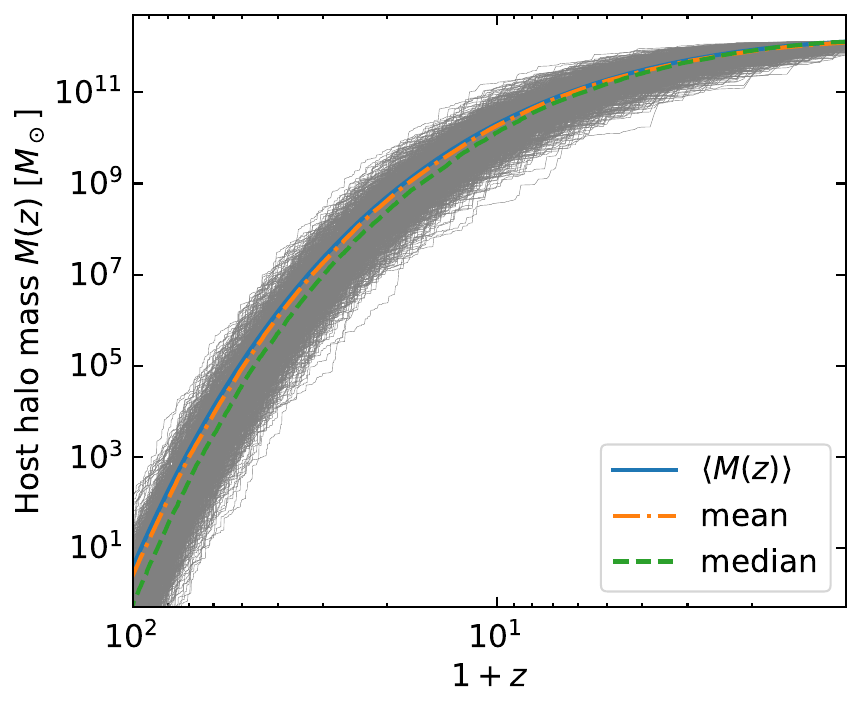}
    \includegraphics[scale=0.45]{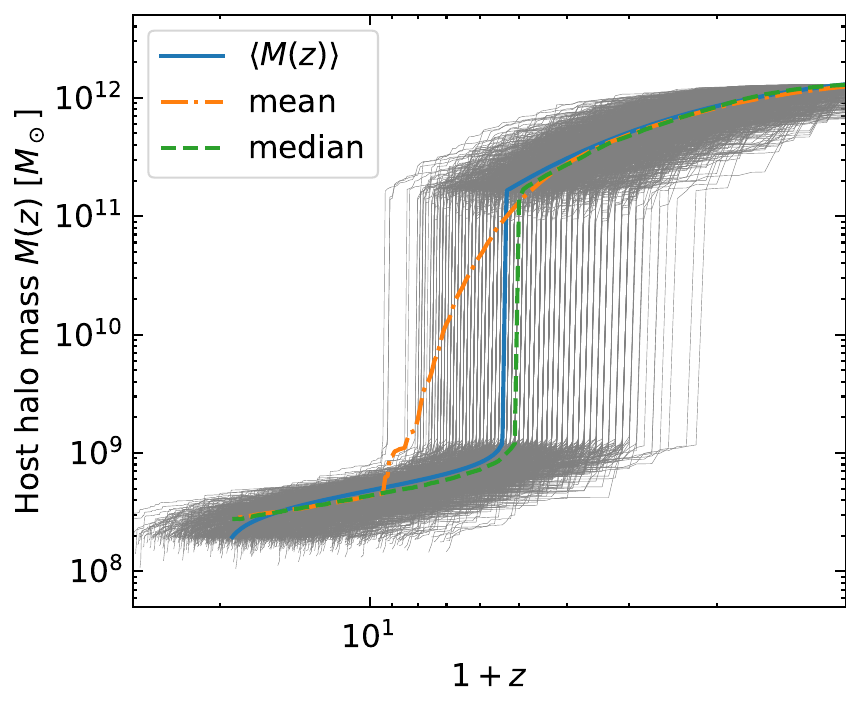}
  \end{center}
  \caption{\small Host halo mass evolution as function of the
    redshift.  The expectation value of host halo mass is plotted in
    thick solid line. Thin gray lines indicate the mass evolution by
    randomly generated number with the PDF, and mean (dot-dashed) and
    median (dashed) values are also shown. (Left) 
    The canonical case, 
    i.e., calculated based on the variance with ${\cal P}_\zeta^{\rm
      CMB}$. (Right) 
      The same parameters as in Fig.\,\ref{fig:Pzeta}
      but with 
    $\Delta=0.1$. }
  \label{fig:Mhost_ran}
\end{figure*}

\begin{figure}[th]
  \begin{center}
    \includegraphics[scale=0.55]{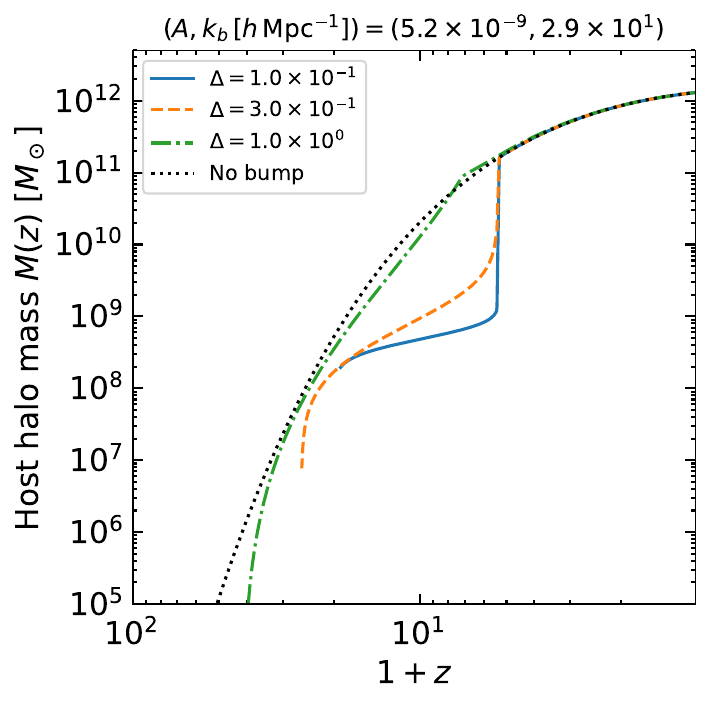}
  \end{center}
  \caption{\small Host halo mass evolution as function of the redshift
    calculated by expectation value. The parameters are the same as
    Fig.\,\ref{fig:Pzeta}. 
    `No bump' shows the result of no bump with
    no cutoff case.}
  \label{fig:Mhost}
\end{figure}

\begin{figure*}[th]
  \begin{center}
    \includegraphics[scale=0.45]{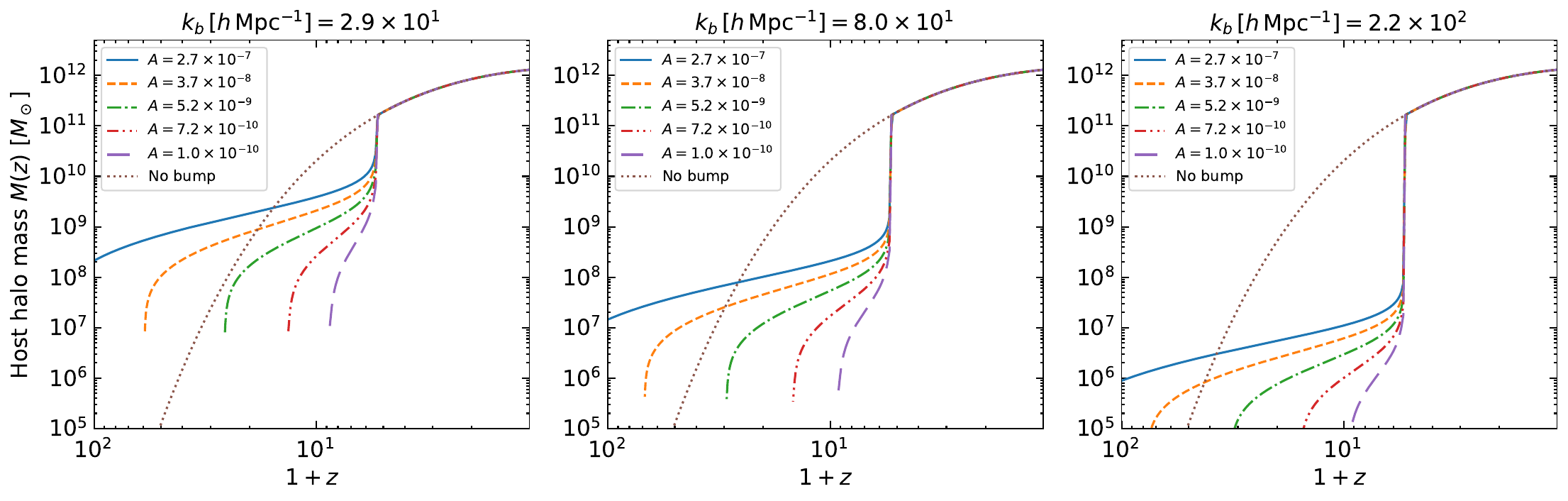}
  \end{center}
  \caption{\small Host halo mass evolution as a function of the redshift calculated by expectation value. The parameters $A$ and $k_b$ are indicated in the figure 
    but with $\Delta=0.1$.
    `No bump' shows the result of no bump with
    no cutoff case.}
  \label{fig:Mhost_delk0.1}
\end{figure*}
In the evaluation of the halo and subhalo evolution, the variance
$\sigma^2$ of the curvature perturbation is the key element, which is
defined by
\begin{align}
  \sigma^2(M) = \int d\ln k \frac{k^3}{2\pi^2}P_M(k)W^2(kR)\,.
\end{align}
$P_M(k)$ is the matter power spectrum and $W(kR)$ is the window
function.  We adopt the sharp-$k$ or $k$-space top-hat window function,
i.e. $W(x)=\Theta(1-x)$, based on a simulation result that reports the
sharp-$k$ window function is suitable for a power spectrum with a steep
cutoff~\cite{Schneider:2013ria}.
We note that the $k$-space top-hat and 
the configuration-space top-hat window functions give almost the same numerical results for the subhalo mass function~\cite{Ando:2022tpj}.
$R$ is the radius that relates to the mass scale $M$ of (sub)halo as
\begin{align}
  M(R)=\frac{4\pi}{3}(R\delta c)^3\rho_M
  \label{eq:M_R}
\end{align}
where $\delta c=2.7$~\cite{Schneider:2013ria} 
is a parameter obtained from a fit to the simulation
and
$\rho_M=\Omega_M\rho_c$ ($\rho_c$ is the critical density) is 
the matter density. 
For
example, $k\,[h,{\rm Mpc}^{-1}]=5$, 40, 30 and 170, correspond to
$M\,[M_\odot]=8.5\times 10^{10}$, $3.9\times 10^{8}$, $2.1\times
10^{7}$, and $2.1\times 10^{6}$, respectively.

Fig.\,\ref{fig:sigma} shows examples of the square root of the variance $\sigma^2(M)$. The high
mass region $\gtrsim 10^{11}\,M_\odot$ corresponds to the integral of
$P^{\rm CMB}_\zeta$ up to the cutoff scale.  An enhancement in the low
mass comes from the contribution of the bump. This behavior gives a
direct consequence to the subhalo mass distribution, which is
discussed in the later section.

\section{Host halo mass evolution}
\label{sec:host}

First, we evaluate the host halo evolution. The growth of the host halo
mass can be modeled by the EPS formalism.  We basically follow the
scheme used in Refs.\,\cite{Ando:2022tpj,Hiroshima:2022khy}, but 
adopt more efficient method to obtain the host halo evolution. In the present study we consider the Milky Way Galaxy as the host halo aiming to probe primordial curvature perturbations using observations of satellite systems of our Galaxy 
and
take its mass as $M_0=1.3\times 10^{12}\,M_\odot$ at $z=0$. However, note that the scheme can be applied to more general situations.

For given host halo mass $M_1$ at $z=z_1$, we ``predict'' the host
halo mass of $M_2$ at $z=z_2=z+\Delta z$ by a probability distribution
function (PDF), which is a function of the variance $\sigma^2(M)$ and
the overdensity $\delta(z)\simeq 1.686/D(z)$.
$D(z)$ is the linear growth factor  
normalized to unity at $z=0$.
Writing
$\sigma^2(M_i)\equiv S_i$ alongside 
$i=1,2$, the PDF to obtain a value of $S_2$ at $z=z_2$ is given by
\begin{align}
  &F_{\rm h}(S_2)\nonumber \\
  &=\left\{
  \begin{array}{ll}
    N_{\rm h}^{-1}f(\delta_2-\delta_1,S_2-S_1) & S_1 \le S_2 \le \sigma^2(M_1/2)
    \\
    0 & \sigma^2(M_1/2)<S_2
  \end{array}
  \right. \,.
\end{align}
where $f$ is the L\'evy distribution:
\begin{align}
  f(\Delta \delta, \Delta s)
  = \frac{\Delta \delta}{\sqrt{2\pi}\Delta s^{3/2}}
  \exp\left[-\frac{\Delta \delta^2}{2\Delta s}\right]
  \,.
\end{align}
Here $N_{\rm h}={\rm erfc}((\delta_2-\delta_1)/\sqrt{2(S_2-S_1)})$ is the
normalization factor where ${\rm erfc}(z)$ is complementary error
function. 
Namely, halos whose masses are larger than $M_1/2$ are  progenitors of the halo of $M_1$ at $z_1$.
The PDF to obtain $M_2$ is given by
\begin{align}
  &\tilde{F}_{\rm h}(M_2)\nonumber \\
  &=
  \left\{
  \begin{array}{ll}
    N_{\rm h}^{-1}
    \left|\dv{S_2}{ M_2}\right|
    f(\delta_2-\delta_1,S_2-S_1)
    & M_1/2<M_2 \le M_1
    \\
    0 & M_2<M_1/2
  \end{array}
  \right. \,.
\end{align}
Using this PDF, we compute the expectation value of $M_2$:
\begin{align}
  \expval{M_2}=\int_{M_1/2}^{M_1}dM \,M\tilde{F}_{\rm h}(M)\,.
  \label{eq:M_exp}
\end{align}
Repeating this procedure at each step of the redshift backwards in time starting from $z=0$, we obtain the
host halo mass evolution. Hereafter we write the result as
$\bar{M}(z)$.  To check the validity
of this procedure, 
we compute the host halo mass
by generating a random number that follows this PDF.

Fig.\,\ref{fig:Mhost_ran} shows the host halo evolution calculated by
the expectation value according to Eq.\,\eqref{eq:M_exp}. In the plot
we also give the result of host halo mass by generating random number
using the PDF $\tilde{F}_{\rm h}$ at each step of the redshift. We
performed this procedure $10^3$ times and obtained the mean and median
values. The left panel shows the result of the canonical case, i.e.,
with neither cutoff nor bump. The redshift dependence of the host halo
mass calculated by the expectation value
is consistent with the mean and and median values obtained from random realizations.
For instance, they coincide at $\order{10}$\,\% level
for $z<5$. We 
also show the case of the power spectrum modeled as Eq.~\eqref{eq:P_zeta_bump}
on the right panel, 
where we take $\Delta=0.1$. 
It shows that $\bar{M}(z)$ well agree with the result of 
the random  
realizations.
Though the mean value deviates from them in the
mass region $[10^{8},10^{11}]~M_\odot$, this is inevitable since the
average cannot describe the discontinuity. 
From the figure, we observe three regimes:
(i) nearly flat, (ii) sharp increase, (iii)
canonical growth, from high to low redshift. This behavior can be understood from the behavior of the 
variance.  Regime (i) corresponds to a sharp drop
in $\sigma(M)$; since the variance drastically decreases around
$10^{8}\,M_{\odot}$, the halo merely grows in this mass range.
Regime
(ii) comes from the flat region $10^{8}\lesssim M\,[M_\odot] \lesssim
10^{11}$ of the variance. In this region, there is no curvature perturbation, 
which results in no production of halos.
Finally the
halo evolution reaches to the canonical track in regime (iii) since
the variance is the same as no bump with no cutoff case.

The results with 
several
values of $\Delta$ are shown in
Fig.\,\ref{fig:Mhost}, where the parameter values are taken to be the same 
as Fig.\,\ref{fig:Pzeta}. With larger value of $\Delta$, the behavior
of the step-like function of $M(z)$ becomes milder. For $\Delta=1$, 
for example, the variance in the region of
$M=[10^8,10^{11}]\,M_\odot$ is not flat any more 
as seen from Fig.~\ref{fig:sigma}. That is why the 
host halo evolves 
similarly as the canonical case when $\Delta$ is large.

We note that the endpoint of the halo evolution line indicates the first halo
formation. Let us denote the endpoint as $(z_{\rm f}, M_{\rm f})$. The halo with the mass $M_{\rm f}$ is first formed at $z=z_{\rm f}$ and there is no halo below $M_{\rm f}$. The mass scale $M_{\rm f}$ and $z_{\rm f}$ depends on $k_b$ and $A$.  Fig.~\ref{fig:Mhost_delk0.1} shows the evolution of the host halo for several choices of $A$ and $k_b$, with $\Delta$ fixed at 0.1. 
$M_{\rm f}$ is determined by $k_b$ with Eq.\,\eqref{eq:M_R},
which can be checked from the figure. On the other hand, $z_{\rm f}$ increases as $A$ becomes larger. This behavior is understood
from the behavior of the variance. As discussed above, the sharp
drop region corresponds to the regime (ii) of the fast halo evolution.  For larger $A$, the sharp 
drop of $\sigma(M)$ is prominent. Thus, regime (ii) gets longer, leading to a large value of $z_{\rm f}$.

To summarize, the unconventional primordial curvature perturbations, i.e., the cutoff and the additional bump, affects
the host halo evolution in the region $z>4$, while the halo 
evolution is the same as the conventional one in the lower redshift
region.  The halo evolution obtained here is consistent with the behavior of the variance and reasonably described.  
Besides, it should be noted that 
as we see in the following sections,
the accretion of subhalo onto the host halo is active only at low $z$ and it is not sensitive to the growth history of the host halo in high $z$.

\begin{figure*}[t]
  \begin{center}
    \includegraphics[scale=0.45]{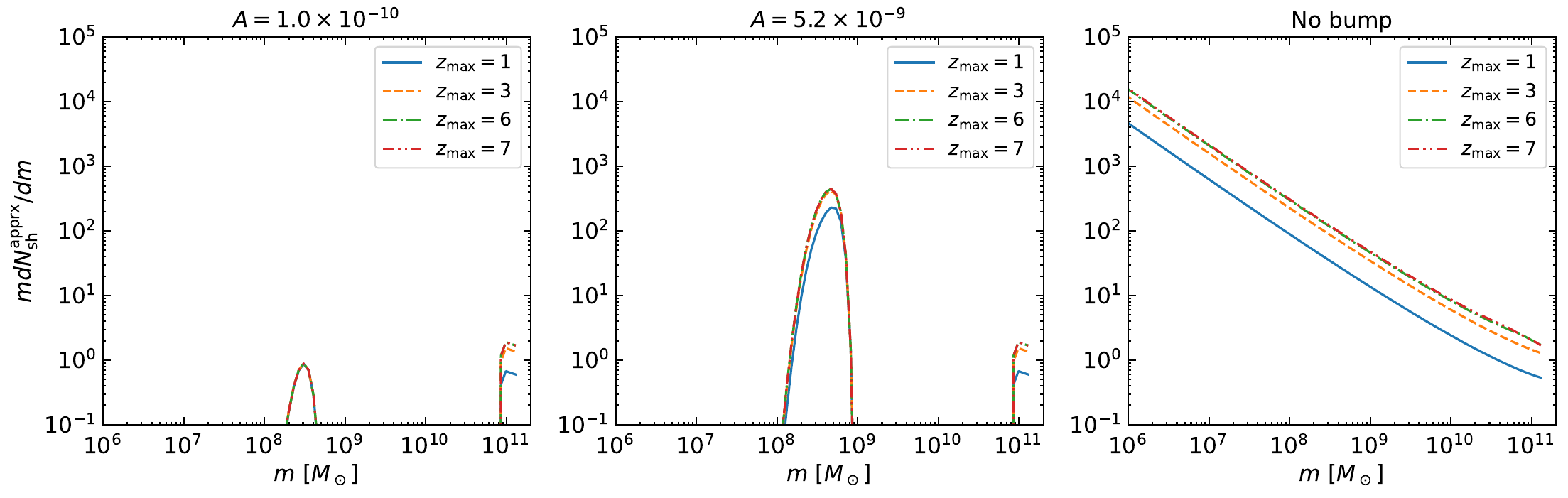}
  \end{center}
  \caption{\small Approximated subhalo mass distribution function $m_a
    dN_{\rm sh}/dm_a$ integrated from 
    $z=0$ to $z=z_{\rm max}$
    as a
    function of subhalo mass.  The left and middle panels are the results
    of the bump with the cutoff model.  $k_b=29\,h$\,Mpc$^{-1}$ and
    $\Delta=0.1$ are taken, and $A=1.0\times 10^{-10}$ (left) and
    $5.2\times 10^{-9}$ (middle). The right panel is the result without
    bump and cutoff. }
  \label{fig:dNdlnm}
\end{figure*}

\section{Subhalo mass distribution}
\label{sec:subhalo}

In the previous section,  we derived the host halo mass evolution in our 
modeling of primordial power spectrum.
Now we discuss its substructure: accretion history of subhalos onto the host halo. We use \texttt{SASHIMI} code~\cite{Hiroshima:2018kfv,Ando:2019xlm} to calculate the subhalo mass distribution function. In order to perform calculations by
applying the numerical package to our model, we incorporate
information on the variance and the evolution of the host halo into
the package. The code is constructed by following the analytical
formulation given in Ref.\,\cite{Yang:2011rf}, including the tidal
stripping effect. The subhalo accretion is based on the EPS
formalism~\cite{Bond:1990iw,Lacey:1993fec}. Here we provide the
essential elements of the EPS theory  
in calculating 
the subhalo mass function, which helps 
interpret the numerical results.

Here we consider a host halo with a mass $M_0$ at the present time and its subhalo
accretion history. In the EPS formalism the number of subhalos with
mass $m_a$ that accretes onto the host halo at the redshift $z_a$ is
given by
\begin{align}
  d^2N^{(0)}_{\rm sh}= \frac{1}{m_a}
  d\bar{M}_a
  {\cal F}_a ds_a\,,
  \label{eq:d2Na_0}
\end{align}
where $s_a\equiv \sigma^2(m_a)$ and $\bar{M}_a\equiv \bar{M}(z_a)$ is the mean host halo mass at $z=z_a$.
$d\bar{M}_a$ is the increment due to the accretion.
Based on the discussion in Sec.\,\ref{sec:host}, we can identify the mean 
and the median values of the host halo mass and
hence use $\bar{M}$ evaluated by Eq.~\eqref{eq:M_exp}.
${\cal F}_a$ describes the
normalized mass distribution of the subhalos that are accreted onto
the host halo. The index ``(0)'' stands for the original, ``unevolved'' subhalo mass
function based on Ref.\,\cite{Yang:2011rf}. To take into account 
possible different paths to the present value of $M_0$, which is depicted as a
bundle of gray lines in Fig.\,\ref{fig:Mhost_ran}, we consider a host
halo of mass $M_a$ at $z_a$. Here $M_a$ is the mass of host halo just
after the accretion, and do not confuse it with $\bar{M}_a$.

${\cal F}_a$ is obtained by averaging a function $F_{\rm s}$ over $M_a$,
\begin{align}
  {\cal F}_a&=\int dM_a P(M_a) F_{\rm s}(s_a,\delta_a;s_M,\delta_M) \,,  
\end{align}
where $P(M_a)$ is the probability distribution function of host halo
mass, which we assume the log-normal distribution with base-10 logarithms
and the dispersion $\sigma$ is given by $\sigma=0.12-0.15 \log
(\bar{M}_a)/M_0$. 
The function $F_{\rm s}$ is the key function to describe
the subhalo mass distribution, which is defined by
\begin{align}
  &F_{\rm s}(s_a,\delta_a;s_M,\delta_M) \nonumber \\
  &=
  \left\{
  \begin{array}{ll}
  N^{-1}_{\rm s} f (\delta_a-\delta_M,s_a-S_M)
 & m_a\le m_{\rm max}
  \\
  0&{\rm otherwise}
  \end{array}
  \right. \,,
\end{align}
where $m_{\rm max}={\rm min}(M_a,M_0/2)$, $\delta_a = \delta (z_a)$,
$\delta_M =\delta (z_M)$, and $S_M \equiv \sigma^2(M_{\rm max})$.
For the argument of $S_M$, we use the Model III of Ref.~\cite{Yang:2011rf}, $M_{\rm max}\equiv {\rm
  min}(\bar{M}_a+m_{\rm max}, M_0)$ instead of $M_a$ since it gives a 
better agreement with the N-body simulation. Then $z_M$ is determined
by $\bar{M}(z_M)=M_{\rm max}$. Finally
$N_{\rm s}=\int^\infty_{\sigma^2(m_{\rm max})}ds_a f(\Delta
\delta,s_a-s_M)={\rm erf}(\Delta \delta/(\sqrt{2\Delta s}))$ is the
normalization factor, where ${\rm erf}(x)$ is error function and
$\Delta \delta = \delta_a-\delta_M$ and $\Delta s = \sigma^2(m_{\rm
  max})-S_M$.

\begin{figure}[th]
  \begin{center}
    \includegraphics[scale=0.55]{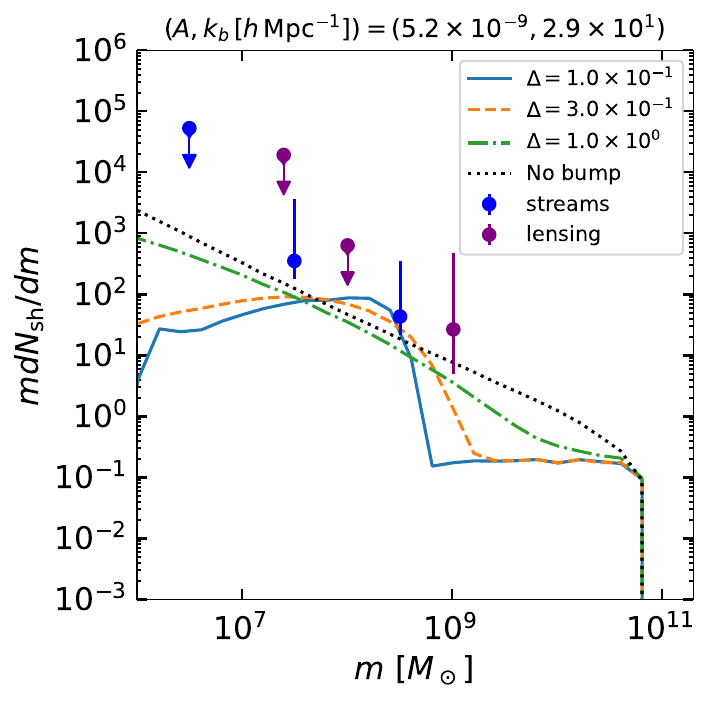}
  \end{center}
  \caption{\small Subhalo mass distribution function calculated by using 
  the tidal stripping
    model~\cite{Hiroshima:2018kfv}. The parameters are the same as
    Fig.\,\ref{fig:Pzeta}. The constraints from stellar streams and
    lensing  measurements~\cite{Hezaveh:2016ltk,Banik:2019cza,Banik:2019smi} are
    shown. 
    }
  \label{fig:SubM_c}
\end{figure}

\begin{figure}[th]
  \begin{center}
    \includegraphics[scale=0.55]{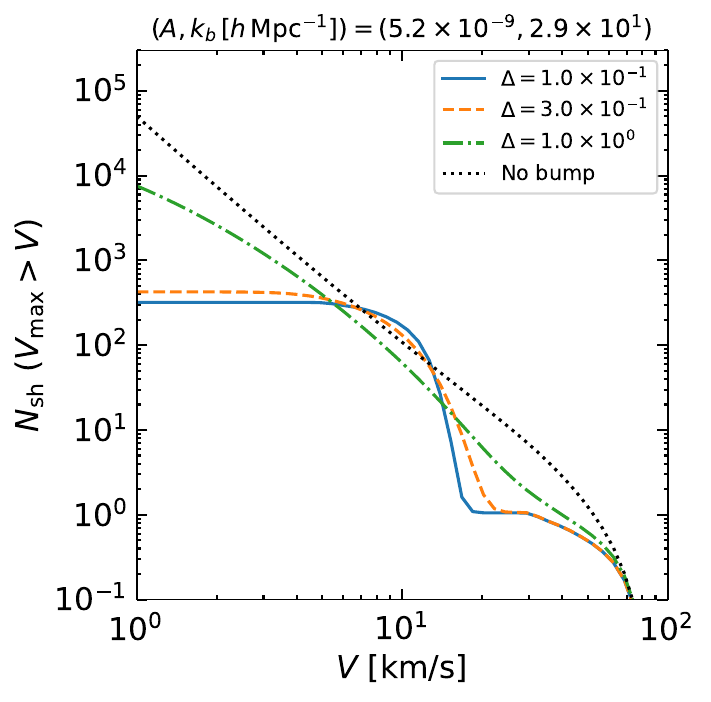}
  \end{center}
  \caption{\small Cumulative maximum circular velocity function
    $N_{\rm sh}(V_{\rm max}>V)$ as a function of $V$. The parameters are
  the same as Fig.\,\ref{fig:Pzeta}.}
  \label{fig:NVmax}
\end{figure}

\begin{figure*}[th]
  \begin{center}
    \includegraphics[scale=0.45]{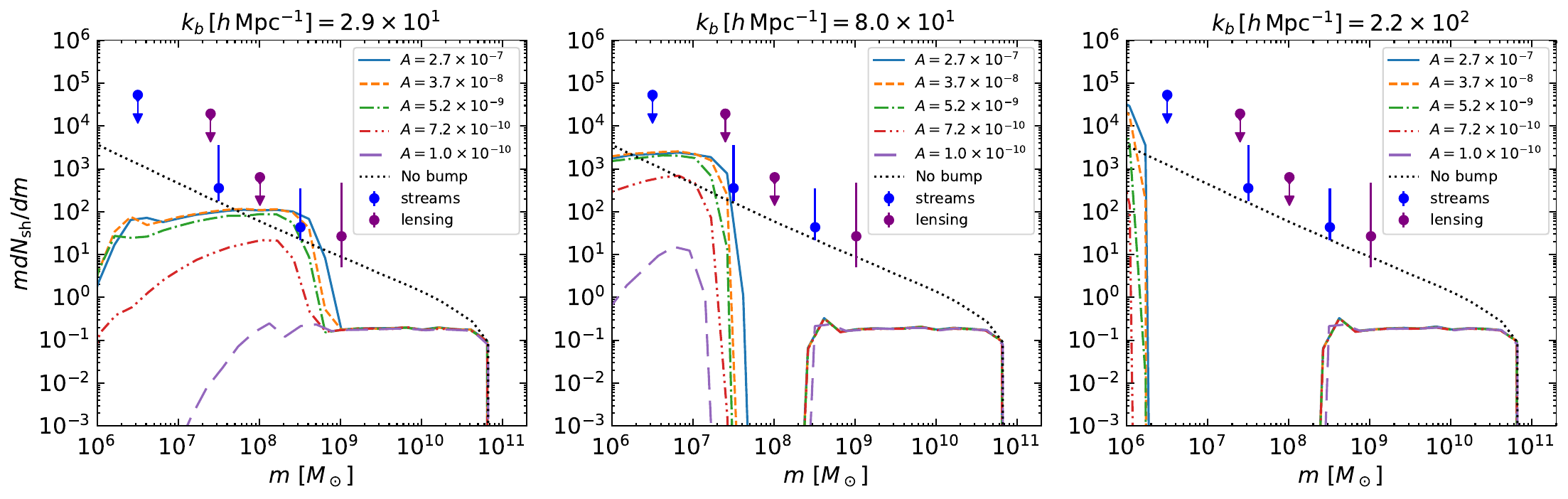}
    \includegraphics[scale=0.45]{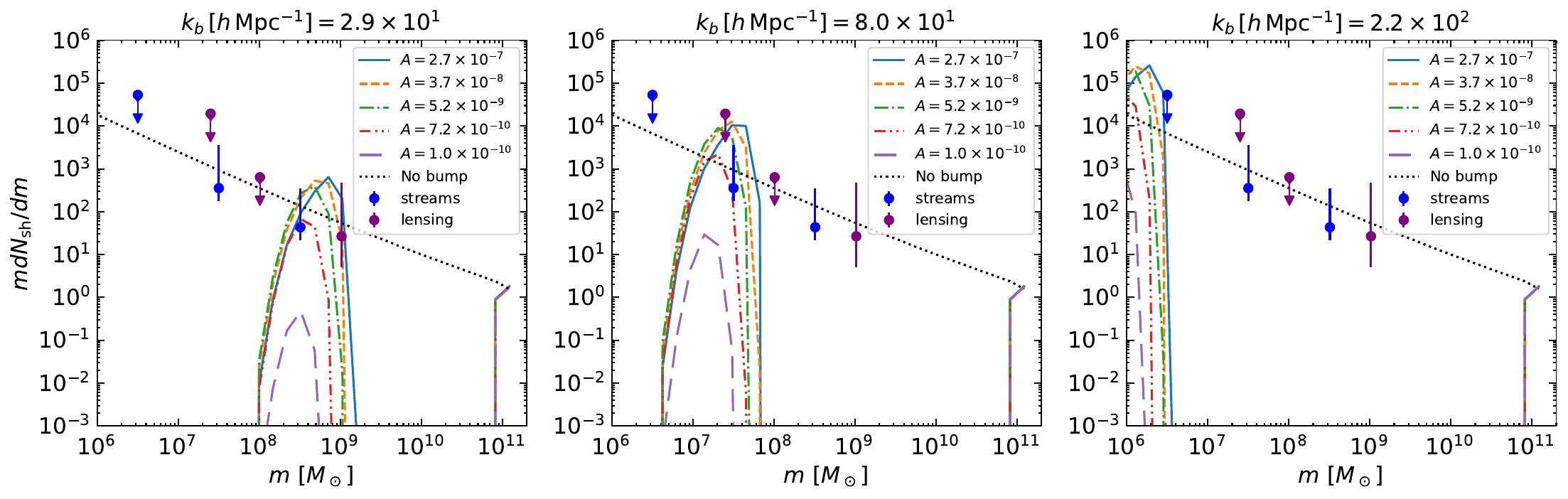}
  \end{center}
  \caption{\small Subhalo mass distribution function calculated by using 
  the tidal stripping model~\cite{Hiroshima:2018kfv} (top) and 
  no tidal stripping (bottom). The parameters are the same as Fig.\,\ref{fig:Mhost_delk0.1}.}
  \label{fig:SubM_c_delk0.1}
\end{figure*}

\begin{figure*}[th]
  \begin{center}
    \includegraphics[scale=0.45]{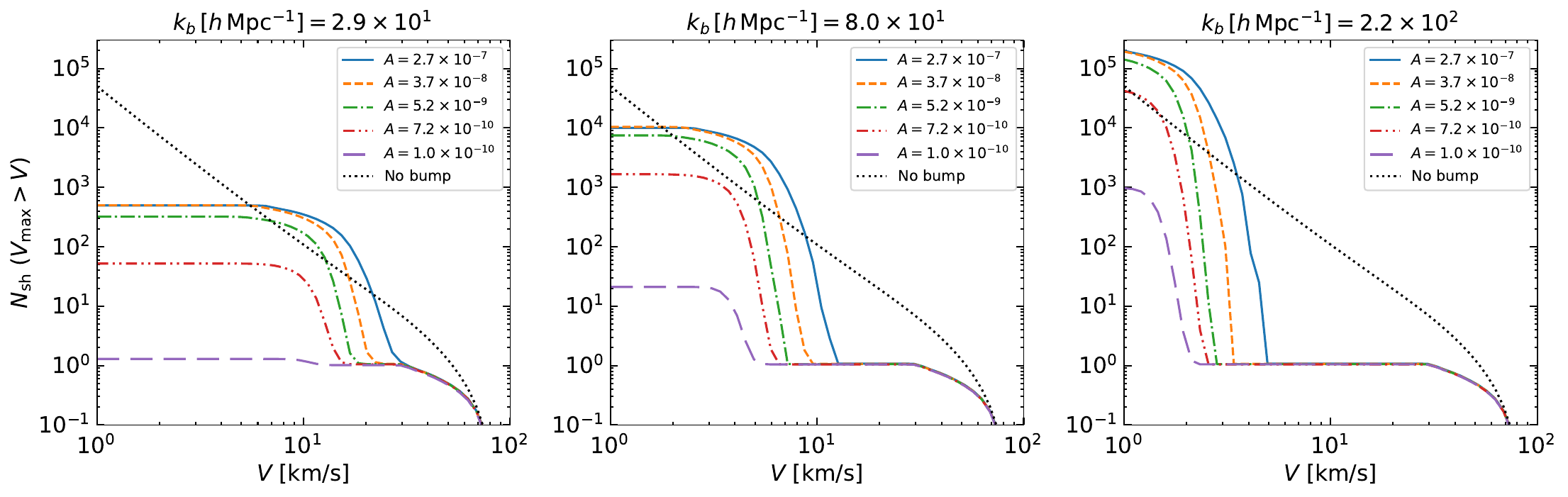}
    \includegraphics[scale=0.45]{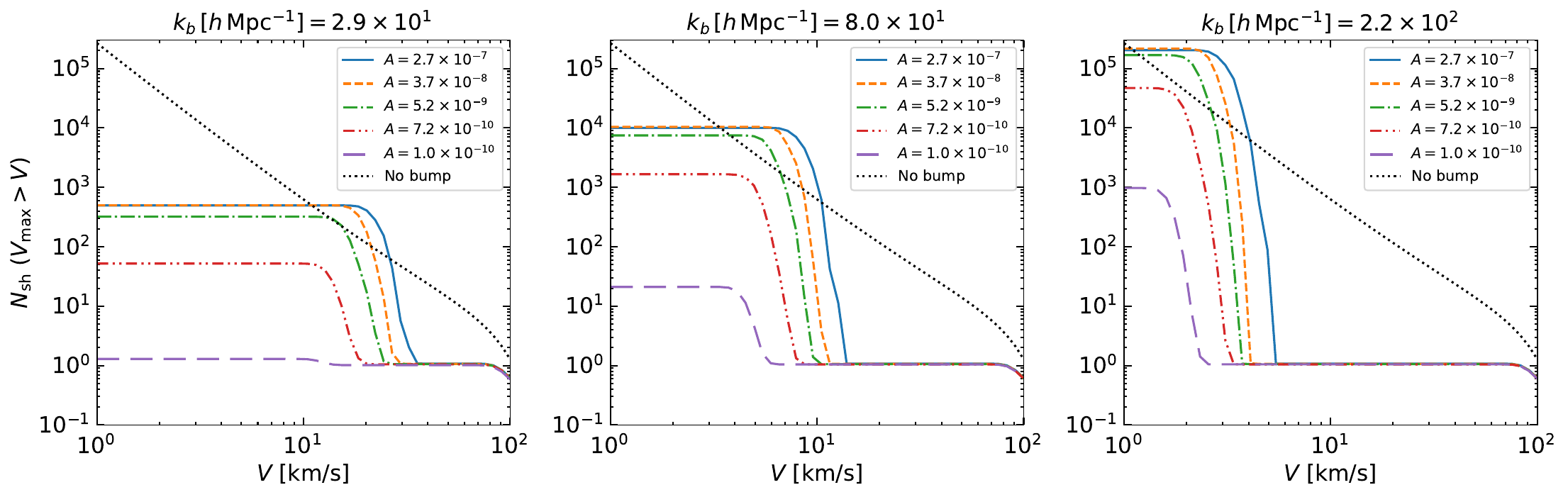}
  \end{center}
  \caption{\small Cumulative maximum circular velocity function
    $N_{\rm sh}(V_{\rm max}>V)$ as a function of $V$. The parameters are
  the same as Fig.\,\ref{fig:Mhost_delk0.1}.}
  \label{fig:NVmax_delk0.1}
\end{figure*}

From Eq.\,\eqref{eq:d2Na_0}, the subhalo mass distribution function at
the redshift $z_a$ without the tidal stripping
effect,
is obtained as
\begin{align}
  m_a\frac{d^2N^{(0)}_{\rm sh}}{dm_adz_a}
  &= {\cal F}_a \frac{ds_a}{dm_a}\frac{d\bar{M}_a}{dz_a}
  \label{eq:d2Nsub_0}
\end{align}
This formula, however, cannot be applied to the present case. To
elaborate the reason, let us make the approximation ${\cal F}_a\simeq
F_{\rm s}$ and replace $M_a$ with $\bar{M}_a$\footnote{
We have checked 
numerically that the approximation 
$M_a \simeq \bar{M}_a$
is good. Compare middle panel of Fig.\,\ref{fig:dNdlnm} and  bottom left panel of Fig.\,\ref{fig:SubM_c_delk0.1}.
}. 
Then the subhalo mass
function is
\begin{align}
  m_a\frac{d^2N^{(0)}_{\rm sh}}{dm_adz_a}
  \simeq F_{\rm s} (s_a,\delta_a;s_M,\delta_M) 
  \frac{ds_a}{dm_a}\frac{d\bar{M}_a}{dz_a}\,.
  \label{eq:SubMfn_app}
\end{align}
Consider the redshift at which the $\bar{M}(z)$ jumps, for instance
$z=4$\,--\,5 for $k_b=80\,h\,{\rm Mpc}^{-1}$. Then, the integral over
the mass region around the bump scale, i.e., $10^7\,M_\odot$, gives
rise to
\begin{align}
  \int dm_a \frac{d^2N^{(0)}_{\rm sh}}{d\ln m_adz_a}
  \sim \frac{d\bar{M}_a}{dz_a}\,.
\end{align}
The right-hand side does not vanish even in the limit of $A\to 0$.
We checked this
behavior numerically without the approximation. However, this is
unphysical because in that limit there should be no halo formed around
the mass scale. 
The problem is attributed to the fact that the variance $\sigma^2$  used in the distribution function $f$ does not vanish even when  $A\to 0$, which also gives a nonzero distribution $f$.
This finite value
arises from the curvature perturbation below the cutoff scale and it
should not be used for the criterion of the collapse of the density
perturbation around the bump scale. To amend this, we introduce the
variance contributed solely by the bump:
\begin{align}
  \sigma^{2}_{\rm bump}(M) = \int d\ln k \frac{k^3}{2\pi^2}P_M^{\rm
    bump}(k)W^2(kR)\,,
\end{align}
where $P_M^{\rm bump}$ is the matter power spectrum 
only taking into account
${\cal P}_\zeta^{\rm bump}$. With the variance 
we define the probability
for the formation of a halo:
\begin{align}
  P_{\rm coll}(z,M)
  =\left\{
  \begin{array}{ll}
    {\rm erfc}\left(\frac{\delta(z)}{\sqrt{2}\sigma_{\rm bump}(M)}\right)
    & M<M_{\rm cut}\\
    1 & M\ge M_{\rm cut}
  \end{array}
  \right. \,,
  \label{eq:Pcoll}
\end{align}
where 
$M_{\rm cut}=M(k^{-1}_{\rm cut})$ 
using Eq.\,\eqref{eq:M_R}. In
our analysis, we multiply this probability function to obtain the
modified subhalo mass function
\begin{align}
m_a\frac{d^2N_{\rm sh}}{dm_adz_a}
&= {\cal F}_a \frac{ds_a}{dm_a}\frac{d\bar{M}_a}{dz_a}
P_{\rm coll}(z_a,m_a)\,.
\label{eq:massfn_mod}
\end{align}
We use this equation to calculate the subhalo mass function for the bump
model. On the other hand, we use Eq.\,\eqref{eq:d2Nsub_0} for the calculation
of the canonical model, i.e., no bump and cutoff in the curvature perturbation.  To compare our results with observational data from stellar streams and gravitational lensing in later
discussion, we apply a correction to the mass function following Ref.\,\cite{Ando:2022tpj}. 
Each point with an error bar of the 
observation, shown in Fig.\,\ref{fig:SubM_c}, for instance,  denotes the subhalo mass function within 300 kpc of the center of the Milky Way halo, as derived in Refs.\,\cite{Hezaveh:2016ltk,Banik:2019cza,Banik:2019smi}.
To convert the subhalo mass function within the virial radius obtained from \texttt{SASHIMI} into that within 300 kpc, we adopt the spatial distribution presented in Fig.\,11 of Ref.\,\cite{Bird:2010mp}.

Fig.\,\ref{fig:dNdlnm}
shows the 
approximated subhalo mass distribution function $m_adN^{\rm
  apprx}_{\rm sh}/dm_a$ where ${\cal F}_a$ and $M_a$ in Eq.\,\eqref{eq:massfn_mod} are replaced by
$F_s$ and $\bar{M}_a$, respectively. Here the integral over the
redshift is 
performed in the range of $z=[0,z_{\rm max}]$.
The figure includes the
results for the bump with cutoff model alongside the canonical case,
i.e., without bump and cutoff. 
For the bump model, we take
$k_b=29$\,$h$\,Mpc$^{-1}$ 
with $\Delta=0.1$.  
The subhalos exist in
limited regions: $m\,[M_\odot]\sim 10^{8}$\,--\,$10^9$ and $10^{11}$. This is
because the derivative of the variance $ds_a/dm_a$ is nearly zero in
the other region. A peak around $m\,[M_\odot]\sim 10^{8}$\,--\,$10^9$ comes from
the bump and the subhalo of $m\,[M_\odot]\sim 10^{11}$ are the
contribution from the curvature perturbation below the cutoff. In the
case of $A=1.0\times 10^{-10}$, the subhalo distribution around
$m\,[M_\odot]\sim 10^{8}$\,--\,$10^9$ is suppressed. This is due to the
probability $P_{\rm coll}$ and the subhalo distribution in that mass
region goes to zero for $A\to 0$, as expected. 

Additionally, we find that the dominant components of the mass
function are in the low $z$ region, i.e., $z\lesssim 5$. This $z$ dependence is the consequence of the derivative of the evolution of the host halo  $d\bar{M}_a/dz_a$. As seen in Fig.\,\ref{fig:Mhost_ran}, $d\bar{M}_a/dz_a$ is suppressed in the region $z>5$. During the period, the host halo evolves slowly; 
that means the accretion of subhalos is
limited. 
That is why the subhalo mass function in the redshift range is
suppressed.  This is also true for the canonical model as seen in the
left panel. In that case, $d\bar{M}_a/dz_a$ become smaller as $z$
gets large. Numerically we find the contributions in $z\gtrsim 6$ is
negligible.

Now we show the subhalo mass distribution function
in Fig.~\ref{fig:SubM_c}
for the same parameters as Fig.\,\ref{fig:Pzeta}.
Additionally, the results for various values of $A$ and $k_b$ with $\Delta$ fixed at $0.1$ are shown in Fig.~\ref{fig:SubM_c_delk0.1}. In this plot, the parameters
are the same as Fig.\,\ref{fig:Mhost_delk0.1}.
Observational data from the gravitational lensing \cite{Hezaveh:2016ltk,Banik:2019smi} and stellar stream measurements \cite{Banik:2019cza,Banik:2019smi} are also shown.
By using the observational data from stellar stream and gravitational lensing measurements, we can derive the limits on the parameters $A$ and $k_b$, which will be presented in the next section.

As mentioned above, the subhalo mass distribution is affected by the tidal stripping. 
To include this effect, 
we use the tidal stripping model given by Ref.\,\cite{Hiroshima:2018kfv}.  We have checked that another model by Jiang \& Bosch~\cite{Jiang:2014nsa}
gives almost the same result. As a reference, we also compute the subhalo mass function without the tidal stripping effect,
which is shown in the bottom panels of Fig.~\ref{fig:SubM_c_delk0.1}.

First let us discuss the result without the tidal stripping. 
The nearly log-normal peak of the mass function arises from the bump in the 
curvature perturbation. 
We observe that the peak value roughly scales as $k_b^{-3}$ for $A$ that is larger than $\order{10^{-8}}$. Since the
peak value corresponds to the number of subhalos of the mass, this
means that the total 
mass
of subhalos, i.e., the integral of the subhalo mass function 
$m dN_{\rm sh}/dm$ over mass, is unchanged.  
This is a somewhat expected result
since the host halo mass evolution is the same and
independent of $k_b$ for $z\lesssim 5$.  Similarly, we can understand
that the peak values almost saturate for $A\gtrsim 10^{-9}$; the total
mass
of subhalos is bound from above due to the host halo mass. On
the other hand, the contribution due to the bump is much smaller for
$A\lesssim \order{10^{-8}}$. This originates in the suppression of the
halo formation probability.

On the other hand, the tidal stripping reduces 
the mass of subhalos 
while leaving their total number unchanged, thereby shifting the subhalo mass distribution toward lower masses.
Due to this effect, the subhalo mass function 
becomes broader, which tends to better fit the observations.

Additionally we comment on the $\Delta$ dependence. 
As can be seen from Fig.~\ref{fig:SubM_c},
when $\Delta$ is as large as unity, the result becomes similar to ``No bump"
case. This is expected since the variance behaves similar to ``No Bump"
shown in Fig.\,\ref{fig:sigma}.

To further check the number of subhalos, we also calculate the cumulative maximum circular velocity function $N_{\rm sh}(V_{\rm max}>V)$: the number of subhalos whose current maximum circular
velocity $V_{\rm max}$ is larger than a value of
$V$. 
Since the circular velocity distribution has a strong correlation with the mass, $N_{\rm sh}(V_{\rm max}>V)$ roughly describes the cumulative mass function of subhalos. 
Fig.\,\ref{fig:NVmax} shows the result 
for the same parameters as Fig.\,\ref{fig:Pzeta}.
 The saturated value corresponds to the total number of subhalos. 
Fig.\,\ref{fig:NVmax_delk0.1} show the dependence on $A$ and $k_b$, where the parameters are the same as Fig.\,\ref{fig:Mhost_delk0.1}.

The cases with and without the tidal stripping effect are also depicted in the figure.
From the figure, it is clear that the saturated value in low-$V$ region, corresponding to the total number of subhalos, is the same with and without the tidal stripping effect.

\section{The constraints from the satellite count, lensing and stellar streams}

\begin{figure*}[t]
  \begin{center}
    \includegraphics[width=5.9cm]{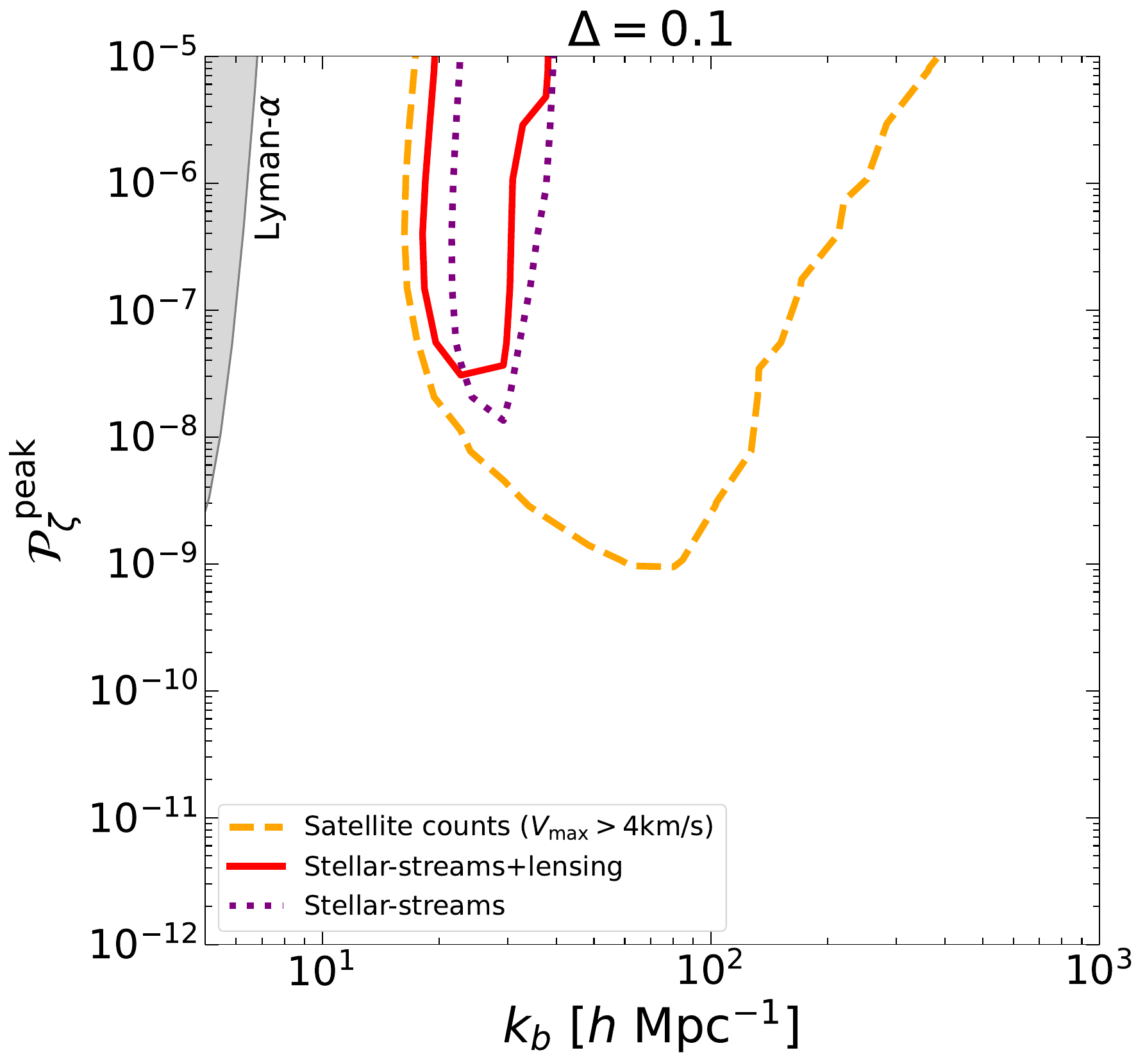}
    \includegraphics[width=5.9cm]{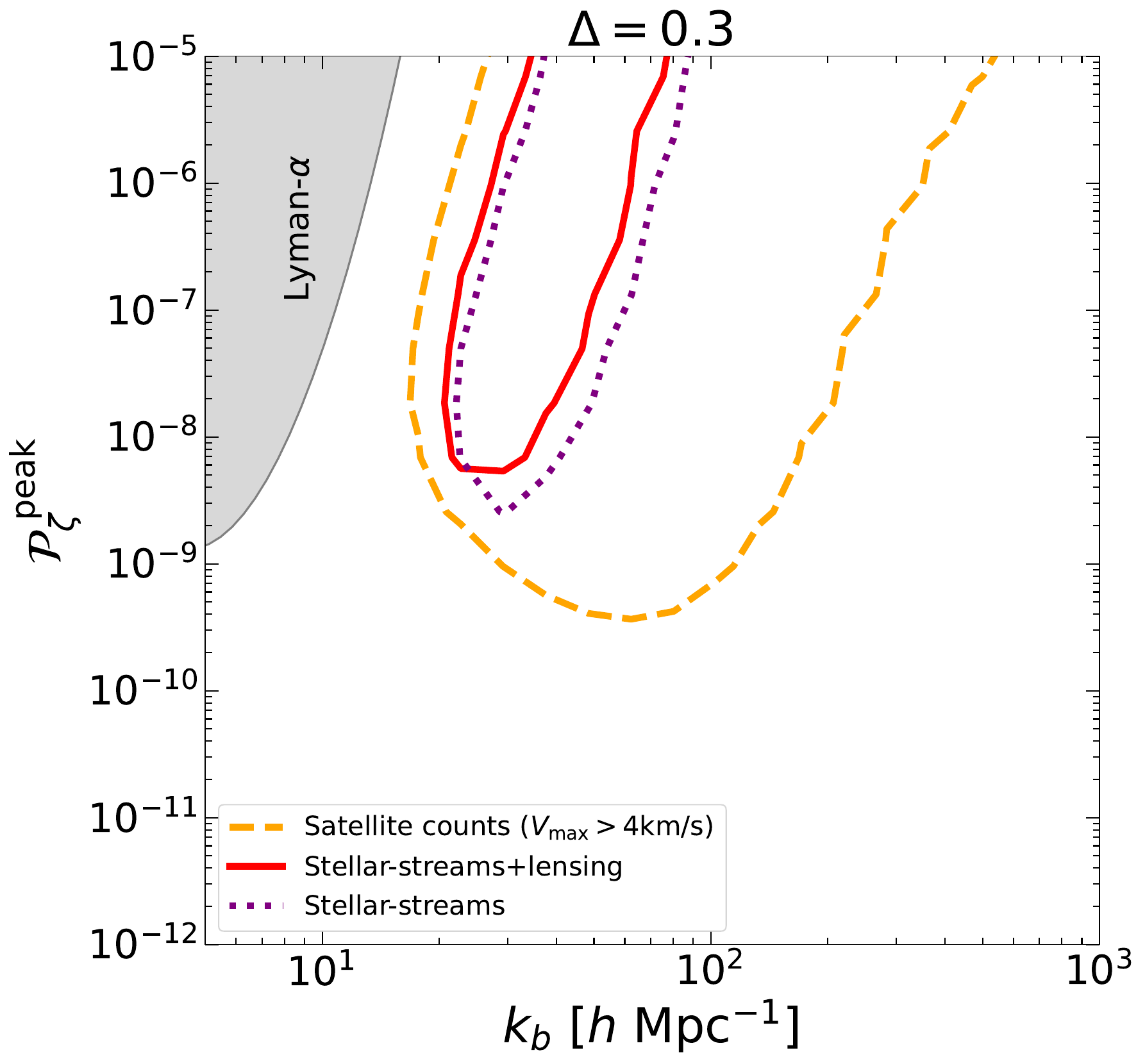}
    \includegraphics[width=5.9cm]{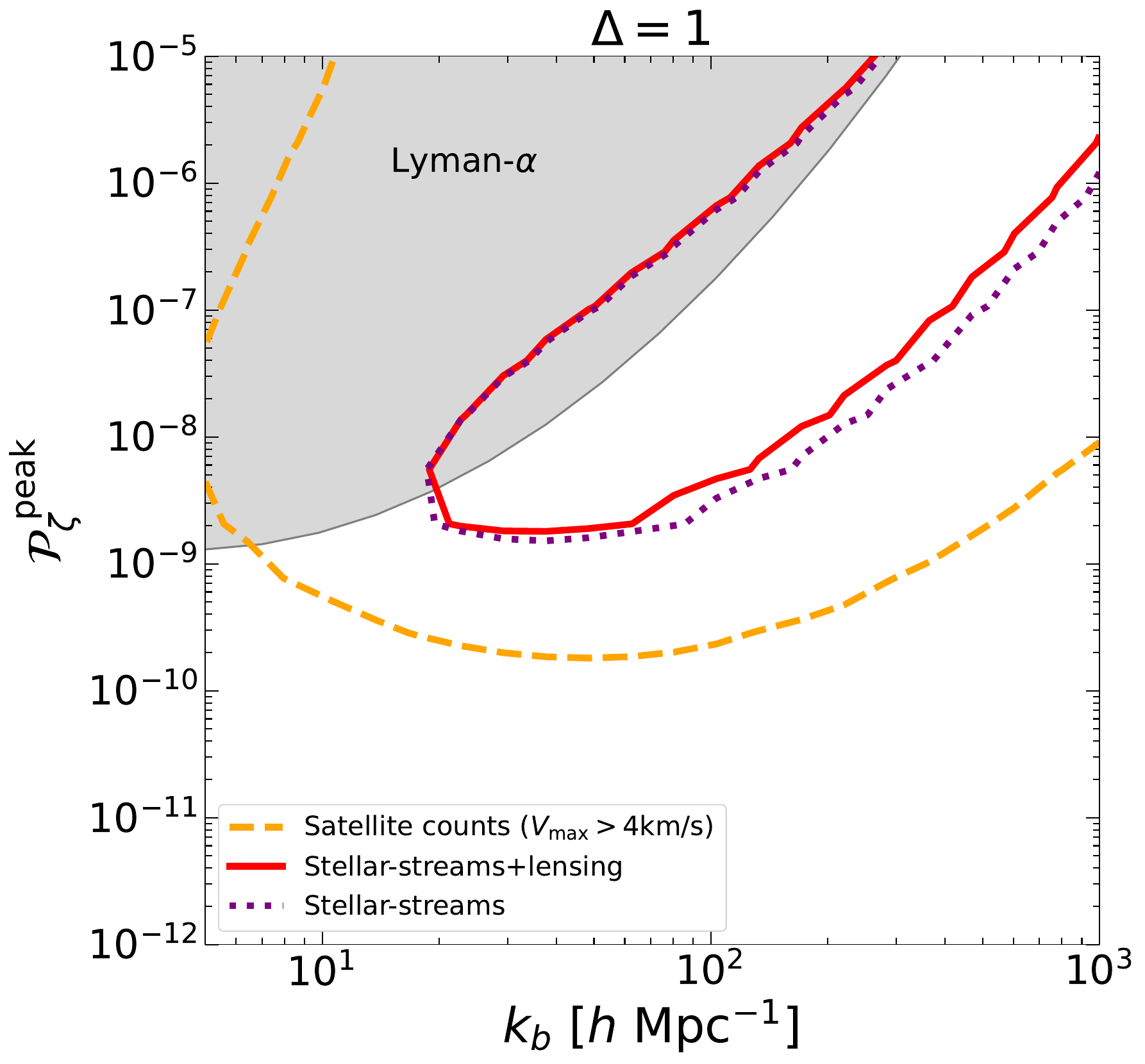}
  \end{center}
  \caption{\small Constraint on $({\cal P}_\zeta^{\rm peak},k_b)$ plane from the gravitational
    lensing, stellar-stream observations and the satellite count. Here ${\cal P}_\zeta^{\rm peak}=A/(\sqrt{2\pi}\Delta)$ is the peak value of the bump in the curvature perturbation.
    Region outside orange
    dashed (red solid) line is excluded by the satellite count
    measurements (lensig and stellar-streams). As a reference, result
    only by using stellar-stream is shown in purple dotted
    line. Shaded region `Lyman-$\alpha$' is excluded by the large
    scale structure measurement~\cite{Bird:2010mp}. } 
  \label{fig:constraint}
\end{figure*}

As seen in the previous sections, the spectrum of the curvature
perturbation has a direct impact on the subhalo mass 
function.
Recent observations are beginning to provide a sharp view thereof.  One useful observable is the number of
dwarf spheroidal (dSph) galaxies of the Milky Way. For these
systems, the present maximum circular velocity, $V_{\rm max}$, is also
measured, and its minimum value is
4~km/s~\cite{Simon:2019nxf,Collins_2017}. The current satellite
count based on the Dark Energy
Survey and PanSTARRS1 survey~\cite{DES:2019vzn,Dekker:2021scf} indicate that there are at least 94
dSphs with $V_{\rm max}>4\,{\rm km/s}$, i.e.,
\begin{align}
  N^{\rm min}_{\rm dSph}(V_{\rm max}>4\,{\rm km/s})=94\,.
\end{align}
In the conventional galaxy formation scenarios, the dSphs 
form
when the maximum circular velocity of subhalo at the accretion exceeds
$18$~km/s, or $10.5$~km/s~\cite{Graus_2019}. To avoid the uncertainty
of the criterion of the galaxy formation, we take a conservative
approach following Ref.\,\cite{Ando:2022tpj}; supposing that all
subhalos become dSphs when $V_{\rm max}>4\,{\rm km/s}$ is satisfied at
the accretion and the number should be larger than 
the one obtained from the observations:
\begin{align}
  N_{\rm sh}(V_{\rm max}>4~{\rm km/s})
  \ge N^{\rm min}_{\rm dSph}(V_{\rm max}>4\,{\rm km/s})\,.
  \label{eq:Nsh_bound}
\end{align}
In our analysis, we derive 95\,\% confidence level (C.L.)
exclusion limit on the model
parameters from this constraint.

Another consequence of subhalos is their impact on gravitational
lensing and stellar streams.  The lensing and stellar stream
observations provide clean probes of subhalos without baryonic
counterparts: they perturb lensed galaxy images and carve gaps in
stellar streams, 
allowing the subhalo abundance in a given mass range to be inferred.
From these signatures, one can 
estimate
their abundance as a function of mass, thereby directly constraining the
subhalo mass distribution function, which is shown in
Fig.~\ref{fig:SubM_c}. With the data, we perform a chi-square
analysis following Ref.\,\cite{Ando:2024ghr}, 
which allows us to derive the constraints on the parameters describing the primordial power spectrum. The chi-square is 
given by
\begin{align}
\chi^2(A,k_b,\Delta)=
\sum_i\frac{[N_i-N_{i,{\rm th}}(A,k_b,\Delta)]^2}
{\sigma_i^2}\,,
\end{align}
where $N_i$ are $N_{i,{\rm th}}$ are the data and
theoretical value of the mass function, respectively, 
and $\sigma_i$ is the 
$1\sigma$ error of the data. $i$ is the label of the data points.
For data points with
error bars, we 
adopt the midpoint of the error bar in logarithmic space as the central value.
For data points with only an upper limit,
we set the central value
to zero and treat the quoted upper limit as the corresponding
2$\sigma$ value. In the present analysis we fix $\Delta$
and give 95\% C.L. 
exclusion limit by 
$\chi^2(A,k_b)-\chi^2_{\rm min}>5.99$.

To present our constraints,
we introduce ${\cal P}_\zeta^{\rm peak}$ as
\begin{align}
{\cal P}_\zeta^{\rm peak} = \frac{A}{\sqrt{2\pi}\Delta}\,,
\end{align}
which is the peak value of the log-normal bump 
in the primordial power spectrum. Then we obtain the
95\% C.L. exclusion limit on $({\cal P}_\zeta^{\rm peak},k_b)$ plane
for a fixed $\Delta$. The result is shown in Fig.\,\ref{fig:constraint} for $\Delta =0.1,\,0.3$ and~$1$. 
In the plot the region outside the curves are excluded. 
Dark grey region is excluded by the large scale structure 
measurement. We adopt the upper bound for the curvature
perturbation ${\cal P}_\zeta<3.1\times 10^{-9}$ at $k=4.5\,h\,{\rm Mpc}^{-1}$~\cite{Bird:2010mp}.
We find that the observation of the gravitational
lensing and stellar streams excludes the region ${\cal P}_\zeta^{\rm peak}\lesssim 10^{-8}/(\Delta/0.1)$. 
As seen in Fig.\,\ref{fig:SubM_c}, 
a bump at $k_b\sim 20$\,--\,40$\,h$\,Mpc$^{-1}$ efficiently produces subhalos in the mass range $10^{7}$\,--\,$10^{9}\,M_\odot$, yielding a subhalo mass function consistent with the observations. Therefore, models that lack sufficient amplitude of the curvature perturbation on these scales are disfavored by the data.

Here are a few remarks on the limit from the gravitational
lensing and stellar streams. For the case with $\Delta=0.1$, the region
consistent with the observation is not sensitive to $A$ if it is larger than $10^{-8}$.
This is because the spectrum of the mass function is
almost saturated at $A\sim \order{10^{-8}}$, which is seen in Fig.\,\ref{fig:SubM_c_delk0.1}.
The allowed region
changes as $\Delta$ increases. When $\Delta$ increases,  
the curvature perturbation at larger wavelength becomes relevant, leading
to abundant larger subhalos. Then, the mass distribution shifts to a 
larger mass region as $A$ gets larger. This is why a bump with a larger
$k_b$ with larger $A$ gives a better fit with the observation. 
We also note that, 
without 
the tidal stripping effect, a large parameter region is excluded for a smaller $\Delta$, such as the case with $\Delta =0.1$ 
since the spectrum of the mass distribution function is too sharp to fit the data. 
This indicates that
the redistribution of the subhalo mass due to the tidal stripping plays an important role to get a better fit with
the observation.  On the other hand, 
a larger value of $\Delta$ broadens the subhalo mass 
function and improves agreement with the observations.

The limit from the satellite count is qualitatively the same for with 
or without the tidal stripping effect.  It gives a lower limit of 
${\cal P}_\zeta^{\rm peak}\sim 10^{-9}/(\Delta/0.1)$, corresponding to 
$A\sim 10^{-10}$. 
The shape of the limit for $\Delta=0.1$ and $0.3$ can be understood qualitatively as
follows. 
The limit in the region $20 \lesssim k_b\,[h\,{\rm
    Mpc}^{-1}]\lesssim 80$ becomes
less severe
as $k_b$ increases.  As
described in Sec.\,\ref{sec:subhalo}, subhalos 
the total mass of subhalos, i.e., the integral of the 
subhalo mass function over mass,
is unchanged for a fixed value of $A$ that is approximately larger than
$P_\zeta^{\rm CMB}$. Writing this as $M_bN_{\rm sh}\sim {\rm const.}$
($M_b$ is the mass scale corresponds to $k_b$), it indicates $N_{\rm
  sh}\propto k_b^{3}$. Then, compiling the knowledge $N_{\rm
  sh}\propto A$, we obtain $N_{\rm sh}\propto Ak_b^3$. Therefore we
expect that the lowest value of $A$ that is consistent with Eq.~\eqref{eq:Nsh_bound} is given by $A \propto k_b^{-3}$. Comparing with
it, the result in Fig.\,\ref{fig:SubM_c_delk0.1}
gives a more
stringent constraint. This is because in this estimation the halo
formation probability given in Eq.\,\eqref{eq:Pcoll} is ignored. Thus,
the actual limit is
more stringent.  In the region 
$k_b\gtrsim 80\,h\,{\rm
  Mpc}^{-1}$, the limit drastically become stringent. The reason is
simple: the larger $k_b$ generates smaller mass of subhalos and it
gets difficult to obtain enough number of $N_{\rm sh}(V_{\rm
  max}>4~{\rm km/s})$. The limit is less stringent if $A$ is larger
and larger because the tail of the bump can generate the subhalos with
larger mass. In the region $k_b\lesssim 20\,h\,{\rm Mpc}^{-1}$, all
region is excluded. This is because the number of subhalos cannot be
larger even if $A$ gets larger.

\section{Conclusion}

We studied the halo and subhalo structure in a given primordial
curvature perturbation in the wavelength that is larger than ${\cal O}(1)$~Mpc$^{-1}$ 
range  
to derive a {\it lower} bound on the amplitude on small scales. 
For this purpose,
we consider that the nearly scale-invariant power spectrum
${\cal P}_\zeta = A_s(k/k_*)^{n_s-1}$ with a cutoff at $k=k_{\rm cut}$ and an
additional log-normal bump in $k>k_{\rm cut}$. 
With this modeling,
we calculate the evolution of Milky-Way like host halos 
and their
subhalos, and then evaluate 
the subhalo abundance for given model parameters such as $A, k_b$ and $\Delta$ to compare with observations of the stellar streams, gravitational lensing, and the satellite counts.

In the calculation of host halo mass evolution, we utilize the
expectation value of halo mass based on the EPS formalism in each step
of the redshift evolution. The result is consistent with the
statistically obtained mean and median values for the stochastically
generated halo mass evolution. 
This procedure significantly reduces the computational cost
since we do not have to generate multiple host
halo mass evolution statistically.

With the result of the host halo mass evolution, we calculate the
subhalo mass function of the host halo. We found that the
observation of the stellar streams
gives a relatively severe {\it lower} bound on primordial power spectrum
in the wavelength of $20$\,--\,$40~h\,{\rm Mpc}^{-1}$.
In particular, when the width of the bump is small, the amplitude should be larger than ${\cal P}_\zeta\sim 10^{-8}$ at the peak.
This is because the perturbation at this
wavelength scale is crucial for the formation of subhalos of
$10^8$\,--\,$10^{10}\,M_\odot$, which corresponds to the region
sensitive to the current stellar stream observation. 
Combining gravitational lensing observations leads to a tighter constraint, although the improvement is modest.

We have also investigated the bound using the satellite count, which gives a less severe limit.
However, the constraint is not sensitive to the location of the bump and allows us to derive a bound over a broad range of scales. In particular, the larger $\Delta$, the broader range of scales can be constrained, in which
the power spectrum becomes similar to the nearly scale invariant one and $k_b$ dependence tends to disappear. On the other hand, for $\Delta<0.1$, the result is almost the same as $\Delta=0.1$.

In this paper, we assumed a specific form for the primordial power spectrum with the bump and the cutoff, modeled as Eq.~\eqref{eq:P_zeta_bump}. Nevertheless, by changing the width and the location of the bump, our results would cover a wide range of models, which would allow us to probe various scenarios via small-scale structure.

\section*{Acknowledgement}

This work was supported in part by the JSPS KAKENHI Grant Numbers JP24K07039 [SA], JP22K14035 [NH],  JP20H05852 [NH], JP26H02044[NH], JP20H01894 [KI], 
JP25K07317 [KI], JP25K01004 [TT],  
JSPS Core-to-Core Program Grant Number JPJSCCA20200002 [KI],
and MEXT KAKENHI Grant Numbers JP23H04515 [TT] and JP25H01543 [TT].
The work of N.H is also supported in part by the MEXT Leading Initiative for Excellent Young Researchers Grant Number 2023L0013.

\bibliography{refs}

@article{Kubota:2022pit,
    author = "Kubota, Mio and Oda, Kin-ya and Rusak, Stanislav and Takahashi, Tomo",
    title = "{Double inflation via non-minimally coupled spectator}",
    eprint = "2202.04869",
    archivePrefix = "arXiv",
    primaryClass = "astro-ph.CO",
    doi = "10.1088/1475-7516/2022/06/016",
    journal = "JCAP",
    volume = "06",
    number = "06",
    pages = "016",
    year = "2022"
}

@article{Bae:2022gkv,
    author = "Bae, Jeong-Myeong and Hong, Sungwook E. and Zoe, Heeseung",
    title = "{Modeling cosmological perturbations of thermal inflation}",
    eprint = "2204.08657",
    archivePrefix = "arXiv",
    primaryClass = "astro-ph.CO",
    doi = "10.1088/1361-6382/ad1214",
    journal = "Class. Quant. Grav.",
    volume = "41",
    number = "1",
    pages = "015024",
    year = "2024"
}

@article{Hong:2015oqa,
    author = "Hong, Sungwook E. and Lee, Hyung-Joo and Lee, Young Jae and Stewart, Ewan D. and Zoe, Heeseung",
    title = "{Effects of thermal inflation on small scale density perturbations}",
    eprint = "1503.08938",
    archivePrefix = "arXiv",
    primaryClass = "astro-ph.CO",
    doi = "10.1088/1475-7516/2015/06/002",
    journal = "JCAP",
    volume = "06",
    pages = "002",
    year = "2015"
}

@article{Starobinsky:1992ts,
    author = "Starobinsky, Alexei A.",
    title = "{Spectrum of adiabatic perturbations in the universe when there are singularities in the inflation potential}",
    journal = "JETP Lett.",
    volume = "55",
    pages = "489--494",
    year = "1992"
}

@article{Hezaveh:2016ltk,
    author = "Hezaveh, Yashar D. and others",
    title = "{Detection of lensing substructure using ALMA observations of the dusty galaxy SDP.81}",
    eprint = "1601.01388",
    archivePrefix = "arXiv",
    primaryClass = "astro-ph.CO",
    doi = "10.3847/0004-637X/823/1/37",
    journal = "Astrophys. J.",
    volume = "823",
    number = "1",
    pages = "37",
    year = "2016"
}

@article{Yokoyama:2000tz,
    author = "Yokoyama, Jun'ichi",
    title = "{Inflation and the dwarf galaxy problem}",
    eprint = "astro-ph/0009127",
    archivePrefix = "arXiv",
    reportNumber = "OU-TAP-142",
    doi = "10.1103/PhysRevD.62.123509",
    journal = "Phys. Rev. D",
    volume = "62",
    pages = "123509",
    year = "2000"
}

@article{Kamionkowski:1999vp,
    author = "Kamionkowski, Marc and Liddle, Andrew R.",
    title = "{The Dearth of halo dwarf galaxies: Is there power on short scales?}",
    eprint = "astro-ph/9911103",
    archivePrefix = "arXiv",
    reportNumber = "CALT-68-2249",
    doi = "10.1103/PhysRevLett.84.4525",
    journal = "Phys. Rev. Lett.",
    volume = "84",
    pages = "4525--4528",
    year = "2000"
}

@article{Enqvist:2019jkb,
    author = "Enqvist, Kari and Sawala, Till and Takahashi, Tomo",
    title = "{Structure formation with two periods of inflation: beyond PLaIn $\Lambda$CDM}",
    eprint = "1905.13580",
    archivePrefix = "arXiv",
    primaryClass = "astro-ph.CO",
    doi = "10.1088/1475-7516/2020/10/053",
    journal = "JCAP",
    volume = "10",
    pages = "053",
    year = "2020"
}

@article{Bringmann:2011ut,
    author = "Bringmann, Torsten and Scott, Pat and Akrami, Yashar",
    title = "{Improved constraints on the primordial power spectrum at small scales from ultracompact minihalos}",
    eprint = "1110.2484",
    archivePrefix = "arXiv",
    primaryClass = "astro-ph.CO",
    doi = "10.1103/PhysRevD.85.125027",
    journal = "Phys. Rev. D",
    volume = "85",
    pages = "125027",
    year = "2012"
}

@article{Byrnes:2018txb,
    author = "Byrnes, Christian T. and Cole, Philippa S. and Patil, Subodh P.",
    title = "{Steepest growth of the power spectrum and primordial black holes}",
    eprint = "1811.11158",
    archivePrefix = "arXiv",
    primaryClass = "astro-ph.CO",
    doi = "10.1088/1475-7516/2019/06/028",
    journal = "JCAP",
    volume = "06",
    pages = "028",
    year = "2019"
}

@article{Chluba:2012we,
    author = "Chluba, Jens and Erickcek, Adrienne L. and Ben-Dayan, Ido",
    title = "{Probing the inflaton: Small-scale power spectrum constraints from measurements of the CMB energy spectrum}",
    eprint = "1203.2681",
    archivePrefix = "arXiv",
    primaryClass = "astro-ph.CO",
    doi = "10.1088/0004-637X/758/2/76",
    journal = "Astrophys. J.",
    volume = "758",
    pages = "76",
    year = "2012"
}

@article{Aghanim:2018eyx,
    author = "Aghanim, N. and others",
    collaboration = "Planck",
    title = "{Planck 2018 results. VI. Cosmological parameters}",
    eprint = "1807.06209",
    archivePrefix = "arXiv",
    primaryClass = "astro-ph.CO",
    doi = "10.1051/0004-6361/201833910",
    journal = "Astron. Astrophys.",
    volume = "641",
    pages = "A6",
    year = "2020"
}

@article{Schneider:2013ria,
    author = "Schneider, Aurel and Smith, Robert E. and Reed, Darren",
    title = "{Halo Mass Function and the Free Streaming Scale}",
    eprint = "1303.0839",
    archivePrefix = "arXiv",
    primaryClass = "astro-ph.CO",
    doi = "10.1093/mnras/stt829",
    journal = "Mon. Not. Roy. Astron. Soc.",
    volume = "433",
    pages = "1573",
    year = "2013"
}

@article{Lacey:1993fec,
    author = "Lacey, Cedric and Cole, Shaun",
    title = "{Merger rates in hierarchical models of galaxy formation}",
    doi = "10.1093/mnras/262.3.627",
    journal = "Mon. Not. Roy. Astron. Soc.",
    volume = "262",
    number = "3",
    pages = "627--649",
    year = "1993"
}

@article{Bond:1990iw,
    author = "Bond, J. R. and Cole, S. and Efstathiou, G. and Kaiser, Nick",
    title = "{Excursion set mass functions for hierarchical Gaussian fluctuations}",
    reportNumber = "CFPA-TH-90-015",
    doi = "10.1086/170520",
    journal = "Astrophys. J.",
    volume = "379",
    pages = "440",
    year = "1991"
}

@article{Yang:2011rf,
    author = "Yang, Xiaohu and Mo, H. J. and Zhang, Youcai and Bosch, Frank C. van den",
    title = "{An analytical model for the accretion of dark matter subhalos}",
    eprint = "1104.1757",
    archivePrefix = "arXiv",
    primaryClass = "astro-ph.CO",
    doi = "10.1088/0004-637X/741/1/13",
    journal = "Astrophys. J.",
    volume = "741",
    pages = "13",
    year = "2011"
}

@article{Hiroshima:2022khy,
    author = "Hiroshima, Nagisa and Ando, Shin'ichiro and Ishiyama, Tomoaki",
    title = "{Semi-analytical frameworks for subhaloes from the smallest to the largest scale}",
    eprint = "2206.01358",
    archivePrefix = "arXiv",
    primaryClass = "astro-ph.CO",
    reportNumber = "UT-HET-137, RIKEN-iTHEMS-Report-22",
    doi = "10.1093/mnras/stac2857",
    journal = "Mon. Not. Roy. Astron. Soc.",
    volume = "517",
    number = "2",
    pages = "2728--2737",
    year = "2022"
}

@article{Ando:2022tpj,
    author = "Ando, Shin'ichiro and Hiroshima, Nagisa and Ishiwata, Koji",
    title = "{Constraining the primordial curvature perturbation using dark matter substructure}",
    eprint = "2207.05747",
    archivePrefix = "arXiv",
    primaryClass = "astro-ph.CO",
    reportNumber = "RIKEN-iTHEMS-Report-22, UT-HET-138, KANAZAWA-22-03",
    doi = "10.1103/PhysRevD.106.103014",
    journal = "Phys. Rev. D",
    volume = "106",
    number = "10",
    pages = "103014",
    year = "2022"
}

@article{Jiang:2014nsa,
    author = "Jiang, Fangzhou and van den Bosch, Frank C.",
    title = "{Statistics of dark matter substructure {\textendash} I. Model and universal fitting functions}",
    eprint = "1403.6827",
    archivePrefix = "arXiv",
    primaryClass = "astro-ph.CO",
    doi = "10.1093/mnras/stw439",
    journal = "Mon. Not. Roy. Astron. Soc.",
    volume = "458",
    number = "3",
    pages = "2848--2869",
    year = "2016"
}

@article{Hiroshima:2018kfv,
    author = "Hiroshima, Nagisa and Ando, Shin'ichiro and Ishiyama, Tomoaki",
    title = "{Modeling evolution of dark matter substructure and annihilation boost}",
    eprint = "1803.07691",
    archivePrefix = "arXiv",
    primaryClass = "astro-ph.CO",
    reportNumber = "KEK-TH-2043",
    doi = "10.1103/PhysRevD.97.123002",
    journal = "Phys. Rev. D",
    volume = "97",
    number = "12",
    pages = "123002",
    year = "2018"
}

@article{Ando:2019xlm,
    author = "Ando, Shin'ichiro and Ishiyama, Tomoaki and Hiroshima, Nagisa",
    title = "{Halo Substructure Boosts to the Signatures of Dark Matter Annihilation}",
    eprint = "1903.11427",
    archivePrefix = "arXiv",
    primaryClass = "astro-ph.CO",
    reportNumber = "HALO-SPECIAL{\_}ISSUE/2020/02",
    doi = "10.3390/galaxies7030068",
    journal = "Galaxies",
    volume = "7",
    number = "3",
    pages = "68",
    year = "2019"
}

@article{Banik:2019cza,
    author = "Banik, Nilanjan and Bovy, Jo and Bertone, Gianfranco and Erkal, Denis and de Boer, T. J. L.",
    title = "{Evidence of a population of dark subhaloes from $Gaia$ and Pan-STARRS observations of the GD-1 stream}",
    eprint = "1911.02662",
    archivePrefix = "arXiv",
    primaryClass = "astro-ph.GA",
    doi = "10.1093/mnras/stab210",
    journal = "Mon. Not. Roy. Astron. Soc.",
    volume = "502",
    number = "2",
    pages = "2364--2380",
    year = "2021"
}

@article{Banik:2019smi,
    author = "Banik, Nilanjan and Bovy, Jo and Bertone, Gianfranco and Erkal, Denis and de Boer, T. J. L.",
    title = "{Novel constraints on the particle nature of dark matter from stellar streams}",
    eprint = "1911.02663",
    archivePrefix = "arXiv",
    primaryClass = "astro-ph.GA",
    doi = "10.1088/1475-7516/2021/10/043",
    journal = "JCAP",
    volume = "10",
    pages = "043",
    year = "2021"
}

@article{Simon:2019nxf,
    author = "Simon, Joshua D.",
    title = "{The Faintest Dwarf Galaxies}",
    eprint = "1901.05465",
    archivePrefix = "arXiv",
    primaryClass = "astro-ph.GA",
    doi = "10.1146/annurev-astro-091918-104453",
    journal = "Ann. Rev. Astron. Astrophys.",
    volume = "57",
    number = "1",
    pages = "375--415",
    year = "2019"
}

@article{Collins_2017,
   title={Dynamical evidence for a strong tidal interaction between the Milky Way and its satellite, Leo V},
   ISSN={1365-2966},
   url={http://dx.doi.org/10.1093/mnras/stx067},
   DOI={10.1093/mnras/stx067},
   journal={Monthly Notices of the Royal Astronomical Society},
   publisher={Oxford University Press (OUP)},
   author={Collins, Michelle L. M. and Tollerud, Erik J. and Sand, David J. and Bonaca, Ana and Willman, Beth and Strader, Jay},
   year={2017},
   month=jan, pages={stx067} }

@article{Dekker:2021scf,
    author = "Dekker, Ariane and Ando, Shin'ichiro and Correa, Camila A. and Ng, Kenny C. Y.",
    title = "{Warm dark matter constraints using Milky~Way satellite observations and subhalo evolution modeling}",
    eprint = "2111.13137",
    archivePrefix = "arXiv",
    primaryClass = "astro-ph.CO",
    doi = "10.1103/PhysRevD.106.123026",
    journal = "Phys. Rev. D",
    volume = "106",
    number = "12",
    pages = "123026",
    year = "2022"
}

@article{DES:2019vzn,
    author = "Drlica-Wagner, A. and others",
    collaboration = "DES",
    title = "{Milky Way Satellite Census. I. The Observational Selection Function for Milky Way Satellites in DES Y3 and Pan-STARRS DR1}",
    eprint = "1912.03302",
    archivePrefix = "arXiv",
    primaryClass = "astro-ph.GA",
    reportNumber = "FERMILAB-PUB-19-604-AE",
    doi = "10.3847/1538-4357/ab7eb9",
    journal = "Astrophys. J.",
    volume = "893",
    pages = "1",
    year = "2020"
}

@article{Graus_2019,
   title={How low does it go? Too few Galactic satellites with standard reionization quenching},
   volume={488},
   ISSN={1365-2966},
   url={http://dx.doi.org/10.1093/mnras/stz1992},
   DOI={10.1093/mnras/stz1992},
   number={4},
   journal={Monthly Notices of the Royal Astronomical Society},
   publisher={Oxford University Press (OUP)},
   author={Graus, Andrew S and Bullock, James S and Kelley, Tyler and Boylan-Kolchin, Michael and Garrison-Kimmel, Shea and Qi, Yuewen},
   year={2019},
   month=jul, pages={4585–4595} }

@article{Ando:2024ghr,
    author = "Ando, Shin'ichiro and Balaji, Shyam and Fairbairn, Malcolm and Hiroshima, Nagisa and Ishiwata, Koji",
    title = "{Supermassive black holes from inflation constrained by dark matter substructure}",
    eprint = "2408.11098",
    archivePrefix = "arXiv",
    primaryClass = "astro-ph.CO",
    doi = "10.1103/ybdn-pr3h",
    journal = "Phys. Rev. D",
    volume = "112",
    number = "10",
    pages = "103034",
    year = "2025"
}

@article{Bird:2010mp,
    author = "Bird, Simeon and Peiris, Hiranya V. and Viel, Matteo and Verde, Licia",
    title = "{Minimally Parametric Power Spectrum Reconstruction from the Lyman-alpha Forest}",
    eprint = "1010.1519",
    archivePrefix = "arXiv",
    primaryClass = "astro-ph.CO",
    doi = "10.1111/j.1365-2966.2011.18245.x",
    journal = "Mon. Not. Roy. Astron. Soc.",
    volume = "413",
    pages = "1717--1728",
    year = "2011"
}
\bibliographystyle{h-physrev5}

\end{document}